# Immobilization of the rare earth fraction in lanthanide phosphates LnPO₄.
# Radiation and hydrolytic resistance of the matrix

D.A. Mikhailov[1], E.A. Potanina[1], V.A. Orlova[2], M.S. Boldin[1], A.A. Murashov[1],

A.V. Voronin[1], A.V. Nokhrin[1], K.K. Kordhenkin[2], D.V. Ryabkov[2], A.I. Orlova[1], V.N. Chuvil'deev[1]

[1] Lobachevskii University of Nizhnii Novgorod, 23 Gagarin Ave., N. Novgorod 603022 Russia

[2] V.G. Khlopin Radium Institute, 28 2nd Murin Dr., St. Petersburg 194021 Russia

e-mail: dmitry.mikhailov@mail.ru

## Abstract

Radiation and hydrolytic resistance of $Eu_{0.054}Gd_{0.014}Y_{0.05}La_{0.111}Ce_{0.2515}Pr_{0.094}Nd_{0.3665}Sm_{0.059}PO_4$ with monazite structure has been studied. The powders were obtained by deposition from solutions. Ceramic specimens with relative density ~97 % were obtained by Spark Plasma Sintering (heating rate $V_h = 50$ ℃/min, sintering temperature $T_s = 1070$ °C, sintering time $t_s = 18$ min). Irradiation by accelerated electrons up to dose of $10^9$ Ge was found not to result in a destruction of the target phase. After the irradiation, an increasing of the leaching rate by an order of magnitude (up to ~$10^{-8}$ g/(cm²·day)) and a reduction of the mechanical properties by 32% as compared to the non-irradiated ceramic specimen were observed.

**Keywords**: enclosing matrix, monazite, Spark Plasma Sintering, radiation resistance, chemical resistance

## Introduction

For long-term storing, all radwaste are transformed into solid materials with high radiation and hydrolytic resistances [1]. Glass, glass ceramics, single-phase and multi-phase ceramics are considered as such materials [2].

Phosphate matrices with various structures have attracted much attention [3-5]. Compounds with monazite structure are proposed for the immobilization of lanthanide and actinide fractions often [3-5]. In particular, phosphates with the monazite structure (natural mineral $CePO_4$ [4]) are promising matrices for the immobilization of the lanthanide and actinide fractions [7, 8]. The natural monazite in the in the metamict state was not found that is probably related to self-annealing of the structure resulting in the crystallization of the irradiated specimens [6].

The radiation and hydrolytic stability of phosphates with monazite structure are studied well enough [7]. However, the authors, as a rule, studied the radiation and hydrolytic stability as the factors (effects) independent on each other. The hydrolytic resistance before irradiation was compared to this one after irradiation in [14]. For the monazite specimen $(La,Cs,Sr,U)PO_4$ irradiated with heavy ions, the cation leaching rates were $\sim 2.6 \cdot 10^{-3}$ $g \cdot mm^{-1} \cdot day^{-1}$ that were close to the values of leaching rate for the non-irradiated specimen ($R_L < 10^{-3}$ $g \cdot m^{-2} \cdot day^{-1}$).

Note that the crystalline compounds with the monazite structure incorporating 1-3 cations, 1-2 cations of which are lanthanides were studied most often [7, 9–13] whereas the RRE fraction has more complex composition. The compounds with the cation composition close to the one of the RRE fraction were studied using compounds with the langbeinite [15], pyrochlore [16], and murataite [17] structures as an example. However, in general, such works were rare.

The goal of the present work was to study the effect of external irradiation with electrons on the hydrolytic stability of the ceramic matrix and on the mechanical properties of this one as well as to compare the matrix properties before and after the irradiation. The object of investigation was the mineral-like $Eu_{0.054}Gd_{0.014}Y_{0.05}La_{0.111}Ce_{0.2515}Pr_{0.094}Nd_{0.3665}Sm_{0.059}PO_4$ ceramic matrix with the monazite structure incorporating the rare earth element (REE) fraction – the product of the radwaste fractioning.

## Experimental

The $Eu_{0.054}Gd_{0.014}Y_{0.05}La_{0.111}Ce_{0.2515}Pr_{0.094}Nd_{0.3665}Sm_{0.059}PO_4$ powder was synthesized by deposition from the aqueous solutions. In this purpose, a stoichiometric quantity of phosphoric acid was added dropwise to the solution of lanthanide nitrates. The resulting gel was mixed during 10 min, then was dried during 12 hrs at 120-140 °C and baked at 350 °C during 2 hrs. The obtained sediment of light green color was annealed in air sequentially at 600 °C (5 hrs), 900 °C (5 hrs), and 1000 °C (5 hrs) without rubbing.

The compaction of the powders into the dense ceramics of 10 mm in diameter and 3 mm in height was performed by Spark Plasma Sintering (SPS) using Dr. Sinter® model SPS-625 setup. Sintering was performed in vacuum (4-6 Pa) by passing millisecond high-power eclectic current pulses through a graphite mold with the powder inside at uniaxial pressure P = 64-80 MPa. SPS was performed by heating the powders with a rate of 100 °C/min up to 560 °C and further up to 1170 °C with a lower rate 50 °C/min. The temperature was measured by Chino® IR-AHS2 pyrometer focused on the outer surface of the graphite mold. The shrinkage of the specimens was measured by a high-precision dilatometer. Holding at the sintering temperature was absent, the total time of the sintering process was ~ 18 min. The cooling down of the specimens from the sintering temperature was performed in a free regime together with the setup.

To carry out the investigations, the ceramic specimen surfaces were polished by diamond pastes using Buehler® Ecomet™ 250 machine.

The densities of the sintered specimens (ρ) were measured by hydrostatic weighting in a distilled water using Sartorius® CPA laboratory balance.

Analysis of the phase composition of the powders and ceramics was carried out by X-Ray Diffraction (XRD) phase analysis using Shimadzu® LabX™ XRD-6000 and Bruker® D8 Advance™ diffractometers (CuK$_\alpha$ filtered radiation).

Functional composition of the compounds after irradiation was studied by Raman Scattering Spectroscopy using Bruker® Senterra™ II Raman spectrometer (lasers with the emission wavelength 532 and 785 nm, spectral range 50-1500 cm$^{-1}$, spectral resolution 1.5 cm$^{-1}$) at room temperature in the frequency range 50 – 1500 cm$^{-1}$.

The microstructure of the specimens was studied using JEOL JSM-6490 Scanning Electron Microscope (SEM) with Oxford Instruments® INCA-350 energy dispersion (EDS) microanalyzer.

The Wickers hardness (H$_V$) was measured using HVS-1000 microhardness tester at the load 0.5 kg (4.9 N). The values of H$_V$ were calculated by averaging over 10 (or more) measurements.

The chemical stability of the ceramics was studied by long-time leaching at room temperature in static regime according to Russian National Standard GOST R 52126-2003. The concentrations of lanthanides in the water samples were measured using ELEMENT™ 2 (Thermo Scientific®, Germany) high-resolution mass spectrometer with induction coupled plasma with an external graduation. The graduation was performed using solutions of ICP-MS-68A-A multi-element standard manufactured by High-Purify Standards (USA).

## Results and Discussion

SEM images of synthesized Eu$_{0.054}$Gd$_{0.014}$Y$_{0.05}$La$_{0.111}$Ce$_{0.2515}$Pr$_{0.094}$Nd$_{0.3665}$Sm$_{0.059}$PO$_4$ powder are presented in Fig. 1. As one can see in the images presented, the particles sizes were 0.2 μm or less. The powder contained agglomerates up to 50-100 μm in sizes.

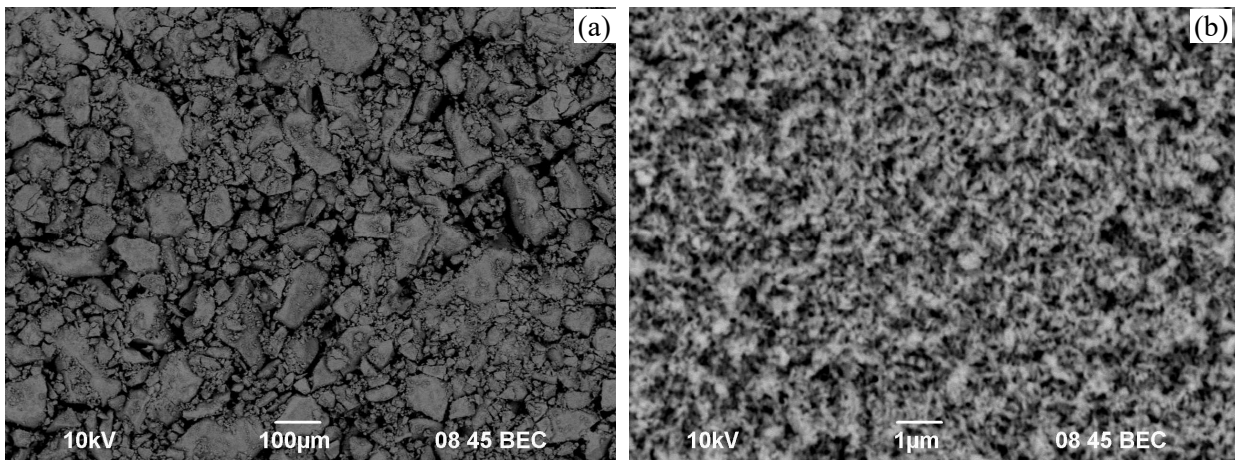

Fig. 1 – SEM images of Eu$_{0.054}$Gd$_{0.014}$Y$_{0.05}$La$_{0.111}$Ce$_{0.2515}$Pr$_{0.094}$Nd$_{0.3665}$Sm$_{0.059}$PO$_4$ powder at different magnifications

According to the XRD phase analysis data (Fig. 2a), the synthesized $Eu_{0.054}Gd_{0.014}Y_{0.05}La_{0.111}Ce_{0.2515}Pr_{0.094}Nd_{0.3665}Sm_{0.059}PO_4$ powder was a single-phase one. The investigated compound crystallized in a monazite-type structure (space group $P2_1/n$, monoclinic syngony, ICSD #25-1065, #83-0654) as expected. The radiographic density of the obtained phase was 5.8323 g/cm$^3$.

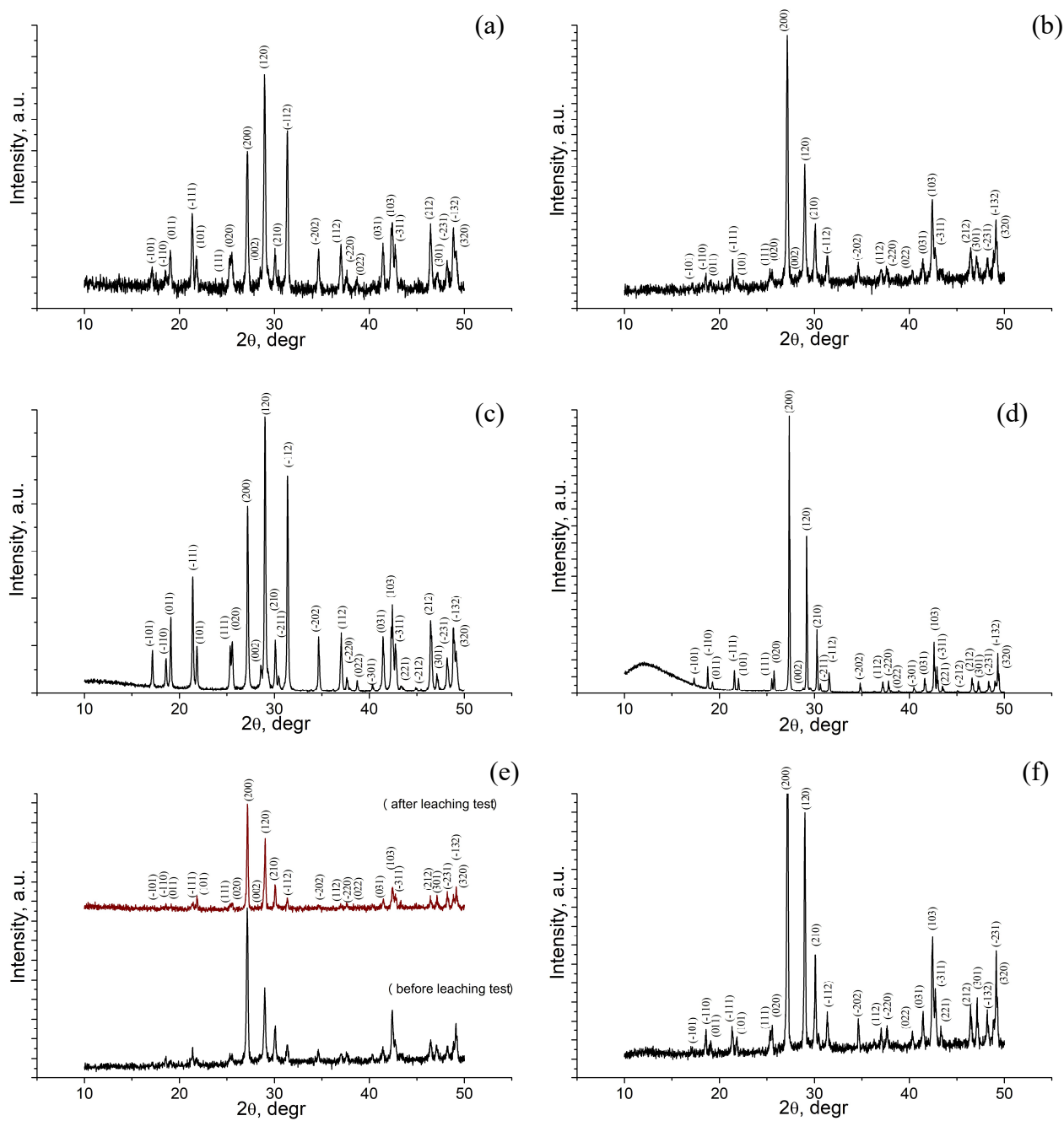

Fig. 2 – Results of XRD phase analysis of $Eu_{0.054}Gd_{0.014}Y_{0.05}La_{0.111}Ce_{0.2515}Pr_{0.094}Nd_{0.3665}Sm_{0.059}PO_4$ compound in the forms of powder (a, c) and ceramic (b, e, f): after irradiation (c, d); after hydrolytic testing (e), after irradiation and hydrolytic testing (f)

The temperature and pressure dynamics during SPS of the ceramic are presented in Fig. 3a. In order to avoid the cracking of the ceramic specimens, a multistage variation of pressure was applied: (I) a gradual increasing of the uniaxial pressure P from 65 MPa up to 79-80 MPa, (II) holding at 79-80 MPa at 580-800 °C; (III) reducing the pressure down to 65 MPa at 900 °C, and again (IV) gradual increasing of the pressure up to ~75 MPa to the moment of the end of the shrinkage. As a result of sintering, dense ceramic specimens without macrodefects were obtained.

Typical dependence of the shrinkage (L) of the powder on the sintering temperature (T) is presented in Fig.. 3b. The analysis of the presented results has shown the dependence L(T) to have a classical three-stage character; the most intensive shrinkage of the powder in SPS takes place in the temperature range 890-1070 °C. The maximum shrinkage rate of the powders was observed at T ~ 1030 °C and was $S_{max} = 2.2 \cdot 10^{-2}$ mm/s.

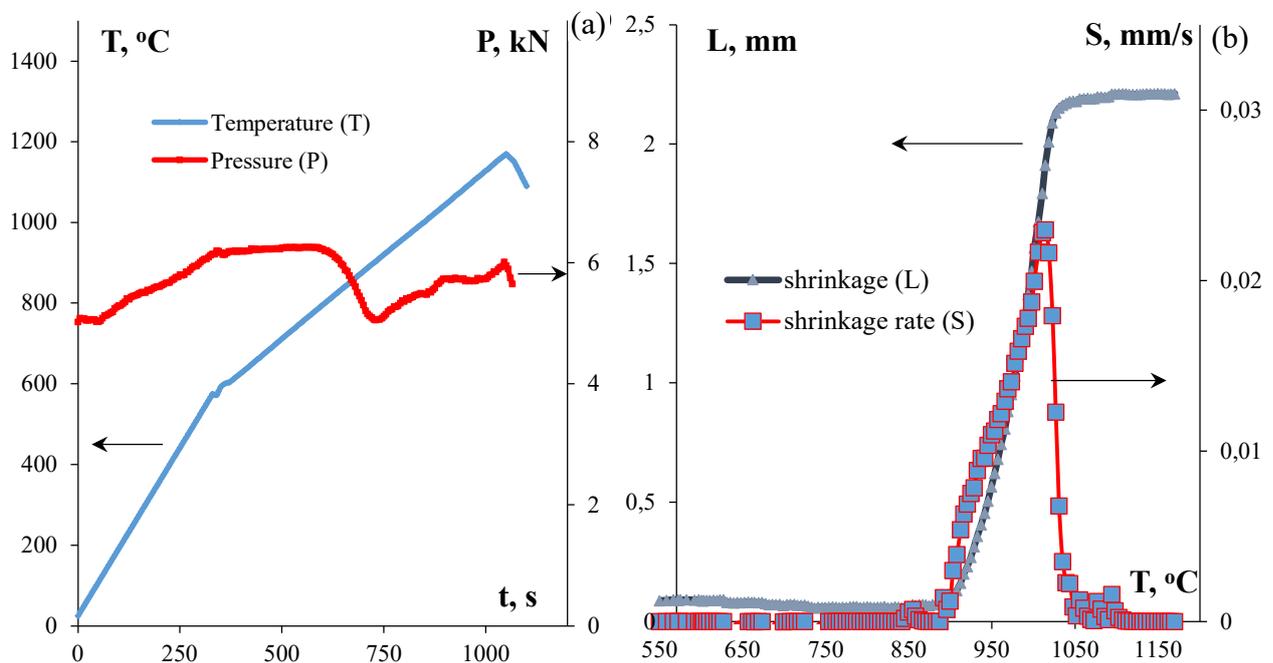

Fig. 3: SPS of the ceramic: a) scheme of the SPS process; b) temperature dependencies of shrinkage (L) and shrinkage rate (S) of the powder during SPS

According to the XRD phase analysis data, the phase composition of the ceramic was identical to the one of the initial powder (Fig. 2b). A comparative analysis of the XRD curves from the powder and of the ones from the ceramic has shown a change in the distribution of the reflection intensities while the positions of these ones remained the same. This effect, in our opinion, may be related to the formation of a texture in the ceramic during sintering under pressure.

The microstructure of the sintered ceramics is presented in Fig. 4a, b. One can see from the SEM images presented that the microstructure of the ceramics was nonuniform. The matrix consisted of the grains of ~ 1.0-5.0 μm in sizes. The abnormal coarse grains of ~15-20 μm in sizes were observed seldom. There were some micrometer- and submicron-sized pores at the grain boundaries. There were no microcracks in the ceramics.

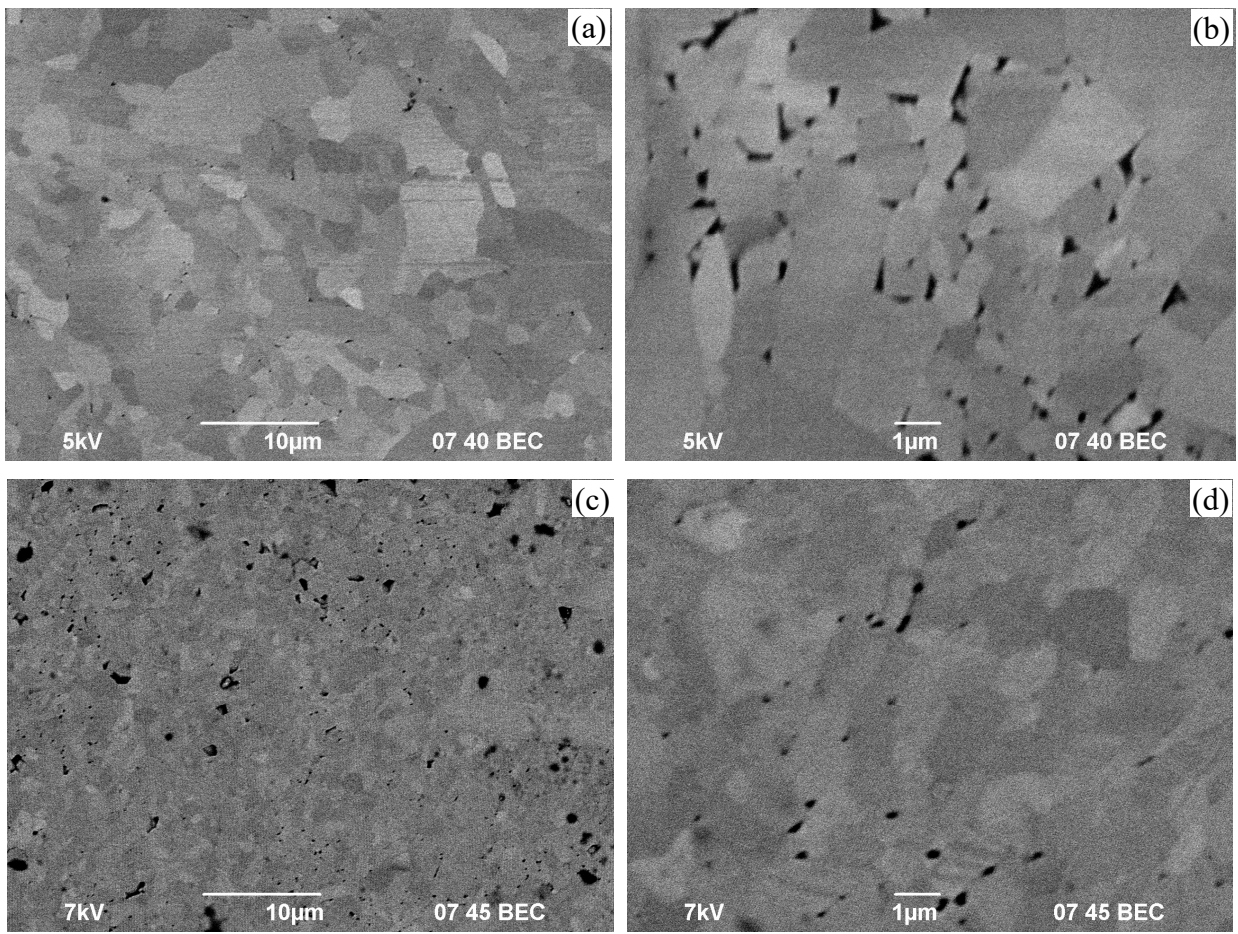

Fig. 4 – Microstructure of the ceramic in the initial state before irradiation (a, b) and after irradiation and leaching (c, d)

The relative density of the sintered ceramic was ~96.6 % (5.634 g/cm$^3$). The averaged microhardness of the ceramic was $H_V = 3.7 \pm 0.3$ GPa.

The phase compositions of the powder and of the ceramic didn't change after electron irradiation with the doses up to $10^9$ Gr (Fig.2c, d). The degree of amorphization after irradiation calculated from the changes in the intensities of the XRD peaks didn't exceed 2%.

According to the Raman spectroscopy data, after the irradiation the crystal structure of the ceramic remained the monazite one (Fig. 5). The series of the monazite bands at 1078, 1053, 1030, and 994 cm$^{-1}$ are related to the antisymmetric valence vibrations $\nu_3$ of the phosphate group $PO_4^{3-}$. The peak at 976 cm$^{-1}$ was related to the symmetric valence vibrations $\nu_1$ of the phosphate group $PO_4^{3-}$. The series of the bands at 622, 592, 572, 561, and 536 cm$^{-1}$ are related to the extraplane bending modes $\nu_4$ of the $PO_4^{3-}$ blocks. The bands at 468, 418, and 398 cm$^{-1}$ were attributed to the bending modes $\nu_2$ in the $PO_4^{3-}$ block planes. The Raman scattering peaks at 279, 262, and 227 cm$^{-1}$ can be related to the metal-oxygen valent bond vibrations. The band series at 177, 157, 131, 122, 104, and 87 cm$^{-1}$ were ascribed to the lattice vibrations related to the presence of various type ions in the monazite structure.

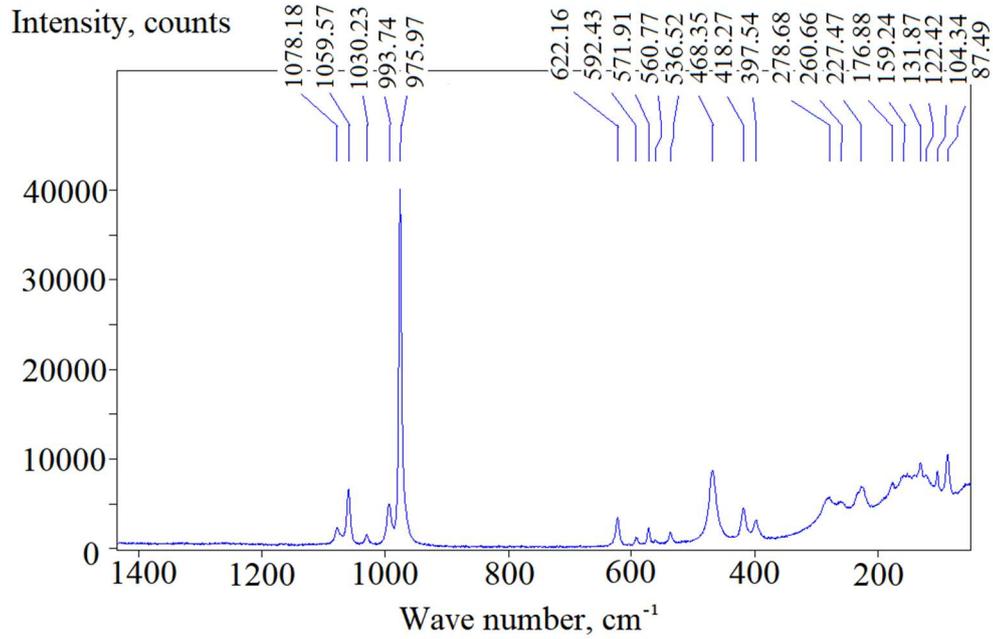

Fig. 5 – Raman scattering spectrum of the ceramic after irradiation

The microhardness of the irradiated ceramic was $2.5 \pm 0.3$ GPa that was less than the one of the non-irradiate ceramic (see above) by ~32%. So far, the irradiation with the accelerated electrons leads to a reduction of the ceramic microhardness without an essential increasing of the volume fraction of the amorphous phase.

The non-irradiated ceramic specimen and the irradiated one were studied for the resistance in the distilled water at room temperature during 28 days. The results of the hydrolytic test are presented in the form of "normalized mass loss – time" and "leaching rate – time" plots in Fig. 6. The minimum cation leaching rates at the 28$^{th}$ day for the non-irradiated ceramic and for the irradiated one are presented in Table 1.

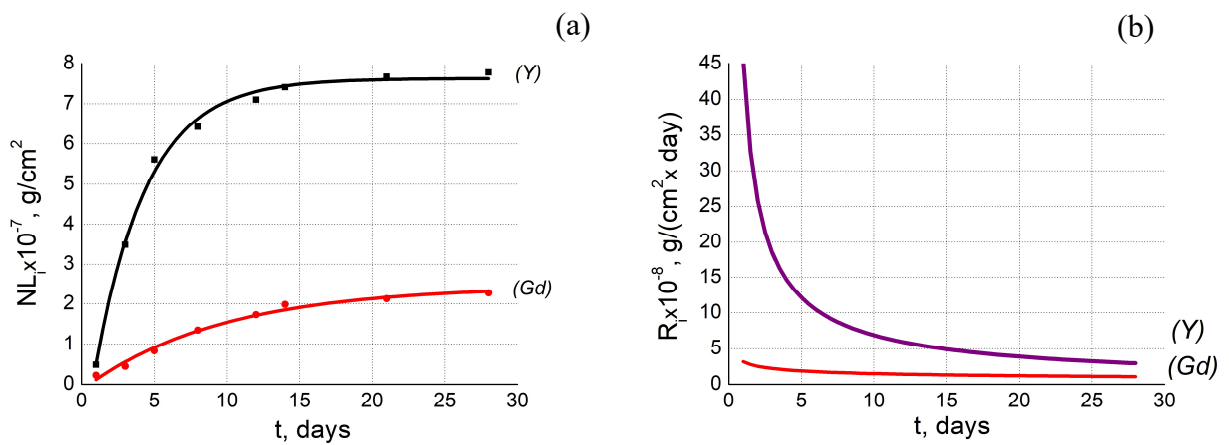

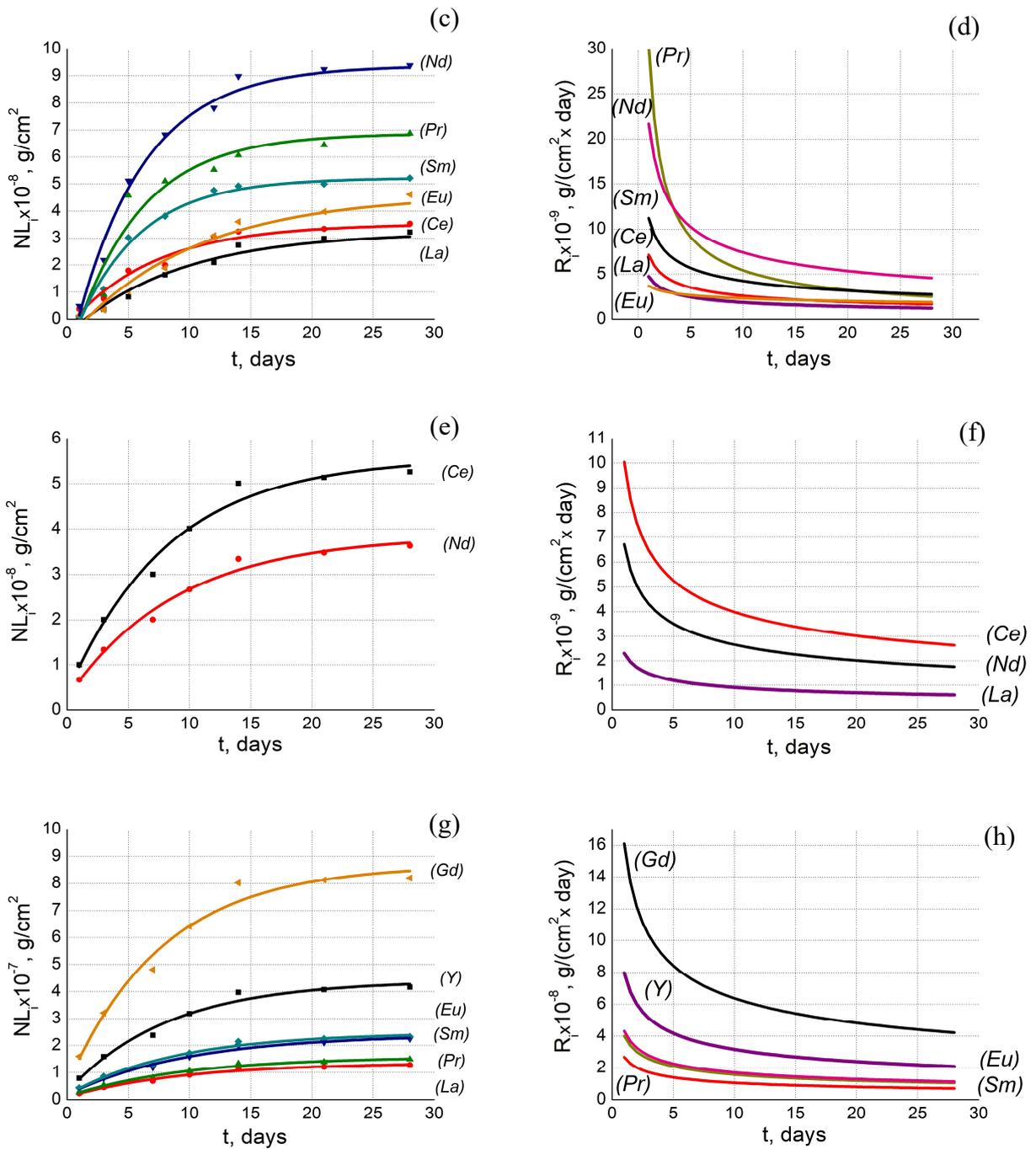

Fig. 6 – Dependencies of the normalized mass losses (NL) and of the leaching rate R on the test time t for the non-irradiated (a-d) and irradiated (e-h) ceramic specimens

Table 1 – Minimum cation leaching rates $R_i$, g·cm⁻²·day⁻¹

| Cation | Eu | Gd | Y | La | Ce | Pr | Nd | Sm |
|---|---|---|---|---|---|---|---|---|
| before irradiation, 10⁹ | 2.0 | 1.2 | 3.7 | 1.4 | 1.9 | 3.1 | 1.8 | 3.1 |
| after irradiation, 10⁹ | 11.3 | 42.1 | 20.9 | 6.0 | 2.6 | 7.0 | 5.2 | 10.5 |

When comparing the data obtained, one can see that the leaching rates of almost all lanthanides increased by an order of magnitude after irradiation. The maximum increasing of the leaching rate $R_i$ after irradiation was observed for Gd (from $1.2 \cdot 10^{-9}$ up to $42.1 \cdot 10^{-9}$ $g \cdot cm^{-2} \cdot day^{-1}$) and for Y (from $3.7 \cdot 10^{-9}$ up to $20.9 \cdot 10^{-9}$ $g \cdot cm^{-2} \cdot day^{-1}$). Note that the irradiation resulted in considerably increased differences in the leaching rates of different cations. In the initial state, the values of leaching rate scattered from 1.2 to 5.2 $g \cdot cm^{-2} \cdot day^{-1}$ whereas after irradiation – from 2.6 to 42.1 $g \cdot cm^{-2} \cdot day^{-1}$. It should be stressed here that in spite of the increasing of the cation leaching rate, the ceramics obtained has a high hydrolytic resistance, in particular, after the irradiation: characteristic cation leaching rates were $\sim 10^{-8}$ $g \cdot cm^{-2} \cdot day^{-1}$.

The analysis of the XRD phase analysis results has shown the ceramic specimens to preserve the original phase composition after hydrolytic testing (Fig. 2e) as well as after the irradiation followed by the hydrolytic test (Fig. 2f).

The microstructure of the ceramic after the electron irradiation and the hydrolytic tests is presented in Figs. 4c, d. One can see the microstructure parameters of the irradiated ceramic correspond to the ones of the ceramic in its original state. The mean grain sizes were $\sim 0.5$-2 μm. A moderate porosity with the pore sizes $\sim 0.1$-1 μm was observed. The damage at the grain boundaries was absent. Comparing the microstructure images of the non-irradiated ceramic (Fig. 4a) and of the irradiated one (Fig. 4c) at the same magnification shows the irradiation to result in an enhanced porosity.

Summarizing the results obtained, one can conclude that $Eu_{0.054}Gd_{0.014}Y_{0.05}La_{0.111}Ce_{0.2515}Pr_{0.094}Nd_{0.3665}Sm_{0.059}PO_4$ ceramic with the monazite structure obtained by SPS is featured by high resistance to the radiation and to hydrolytic impacts.

## Conclusions

The $Eu_{0.054}Gd_{0.014}Y_{0.05}La_{0.111}Ce_{0.2515}Pr_{0.094}Nd_{0.3665}Sm_{0.059}PO_4$, phosphates playing the role of a consolidating matrix for the REE fraction was synthesized by deposition from a solution. As a result of synthesis, the agglomerated submicron-grained powders were obtained. The ceramic with the relative density $\sim 97\%$ was obtained by Spark Plasma Sintering at 1070 °C; the total sintering process duration was 18 min.

The resistance of the ceramic to the radiation and hydrolytic impacts, in particular, the hydrolytic resistance after irradiation has been studied. No phase transformations in the material were observed after electron irradiation with the doses up to $10^9$ Gr. The volume fraction of the amorphous phase didn't exceed 2%. The lanthanide leaching rates in the static regime at room temperature after the irradiation increased by an order of magnitude and reached $\sim 10^{-8}$ $g \cdot cm^{-2} \cdot day^{-1}$ that allows classifying the material obtained as the radiation and hydrolytically proof ones.

The irradiation resulted in a reduction of the microhardness of the ceramic by 32% as compared to the non-irradiated one.

Acknowledgements

The authors are grateful to V.N. Lomasov (Peter the Great Saint-Petersburg Polytechnic University) for carrying out the radiation tests.

Funding

This work has been performed with the support of Research and Educational Center of Nizhny Novgorod region in the framework of the Agreement No. 16-11-2021/49 and in the framework of Contract No 217/4571-D "Substantiation of the radiation resistance of the KP matrices. Support of testing the technologies".

Conflict of Interests

The authors declare no conflict of interests.